\documentclass[prl,twocolumn,showpacs]{revtex4}
\usepackage{amssymb}

\usepackage{graphicx,amsmath}

\newcommand{\bea}{\begin{eqnarray}}
\newcommand{\eea}{\end{eqnarray}}
\newcommand{\ben}{\begin{equation}}
\newcommand{\een}{\end{equation}}
\newcommand{\benu}{\begin{enumerate}}
\newcommand{\enu}{\end{enumerate}}

\newcommand{\tg}{\gamma}

\begin{document}

\title{Quantum Correction to Conductivity Close to Ferromagnetic 
Quantum Critical Point in Two Dimensions}
\date{\today}
\author{I. Paul$^1$, C. P\'epin$^1$, B. N. Narozhny$^2$, and D. L. Maslov$^3$}

\affiliation{$^1$SPhT, CEA-Saclay, L'Orme des Merisiers, 91191
Gif-sur-Yvette, France\\
$^2$The Abdus Salam ICTP, Strada Costiera 11, Trieste, 34100, Italy\\
$^3$Dept. of Physics, University of Florida, P. O. Box 118440,
Gainesville, FL 32611-8440, USA} 

\begin{abstract}
We study the temperature dependence of the conductivity due to quantum
interference processes for a two-dimensional disordered itinerant electron
system close to a ferromagnetic quantum critical point.
Near the quantum critical point, the cross-over between diffusive and
ballistic regimes of quantum interference effects occurs at a temperature $%
T^{\ast }=1/\tau \gamma (E_{F}\tau )^{2}$, where $\gamma $ is the parameter
associated with the Landau damping of the spin fluctuations, $\tau $ is the
impurity scattering time, and $E_{F}$ is the Fermi energy. For a generic
choice of parameters, $T^{\ast }$ is smaller than the nominal crossover
scale $1/\tau $. In the ballistic 
quantum critical regime, the conductivity behaves as $T^{1/3}$.
\end{abstract}
\maketitle

The interplay between disorder, electron correlations, and low
dimensionality is one of the most fascinating topics in the modern condensed
matter. To date, most of the studies were limited to the case of ``good
metals'' which, at high enough temperatures, behave as Fermi Liquids (FL) ~
\cite{aleiner1,zna,altshuler2}. However, this interplay is expected to
become crucial in the vicinity of a quantum critical point (QCP)
where electron correlations are particularly strong~\cite{stewart,julian}.
Experiments on systems close to quantum phase transitions show striking
deviations from the FL theory. In particular, anomalous exponents in the
temperature dependence of the conductivity have been observed~\cite
{uge2,zrzn2},
which suggest the presence of strong quantum fluctuations. Of special
interest is the case of charge transport in the vicinity of a ferromagnetic
QCP. Since ferromagnetic spin fluctuations do not break any lattice
symmetry, the contribution of inelastic scattering to resistivity is zero in
a clean system, unless Umklapp processes are allowed to relax momentum. In a
dirty system,
the ``interaction'' correction to the residual conductivity is expected to
be particularly important due to a long-range interaction in the vicinity of
the QCP. This correction is due to quantum interference between
semi-classical electron paths scattered by the impurities and the
self-consistent potential of Friedel oscillations~\cite{aleiner1}. 
The goal of this paper is to examine the conductivity of a
two-dimensional (2D)
 disordered
metal close to a ferromagnetic QCP and at low enough temperatures, when the
lattice-mediated scattering at spin fluctuations is frozen out and the
temperature dependence of the conductivity is mainly due to quantum
interference effects. 

The experiments indicate that most of the three-dimensional compounds, such
as UGe$_{2}$~\cite{uge2} and ZrZn$_{2}$~\cite{zrzn2}, undergo a first-order
zero-temperature ferromagnetic transition. More recently, the transition
observed in Zr$_{1-x}$Nb$_{x}$Zn$_{2}$ is found to be second order down to
the lowest measured transition temperature~\cite{aronson}. 
In two dimensions, the best candidate for a ferromagnetic type of
quantum critical behavior is the metamagnetic transition
in Sr$_{3}$Ru$_{2}$O$_{7}$~\cite{perry,grigera}. 
This strongly anisotropic compound 
can be tuned to a quantum critical end point which is believed to be
suitable for a description within the spin fluctuation scenario~\cite
{schofield}. Transport properties of disordered metallic systems are well
understood in the case when the electron-electron interaction is weak enough
(so that the system is away from any QCP and the symmetries of the FL state
are not broken)~\cite{altshuler2,lee}. At low enough temperature ($T$), the $%
T$-dependence of the conductivity (as well as other transport coefficients)
is mostly due to quantum interference \cite{altshuler2}. The effect is more
dramatic in lower dimensions, where the temperature dependent corrections to
the residual conductivity exhibit singular behavior. In particular, in 2D
the corrections are logarithmic in the diffusive regime, when $T\tau \ll 1$~
\cite{altshuler2}, and linear in the ballistic regime $T\tau \gg 1$~\cite
{zna}, where $\tau $ is the elastic scattering lifetime of the electrons.
Quantum correction to conductivity has also been studied in the context of
fermion gauge field models\cite{mirlin,khveshchenko}.

Near a QCP the interaction between electrons is strong,
making it difficult to formulate a controlled theory. Therefore it is not
surprising that there has been very few studies of transport properties near
quantum criticality~\cite{kim,rosch1,belitz2}. For a metamagnetic QCP in 2D
it has been shown~\cite{kim} that the conductivity in the diffusive regime
behaves as $\ln ^{2}T$, in contrast to the usual logarithmic temperature
dependence in a good metal~\cite{altshuler2}.

In this Letter we study the conductivity ($\sigma $) of a disordered 2D
metal near a ferromagnetic QCP, assuming the system to be in a continuum
where lattice effects are absent. In the conventional approach to 
QCP~\cite{hmm}, the conduction electrons are integrated out, and
a generalized Landau-Ginzburg action in terms of the order parameter fields
is studied. Recently, the validity of integrating out low-energy electrons
has been questioned~\cite{abanov,belitz1}, and it has been argued that such
an approach generates singularities to all orders in the collective spin
interactions. Here, we start with the phenomenological spin-fermion model of
Ref.~\cite{chubukov1}, which describes the low-energy properties of
electrons close to a ferromagnetic instability, and add scattering of
electrons due to static impurities. For completeness, we also take into
account the coupling to the long-range Coulomb interaction in the singlet
(charge) channel, so that the total correction to the conductivity is the
sum of singlet and triplet contributions: $\delta \sigma (T)=\delta \sigma
_{S}(T)+\delta \sigma _{T}(T)$. Both in the diffusive and ballistic regimes,
$\delta \sigma _{S}(T)$ has an insulating-like behavior common to all
metals~\cite{zna}, which competes with the metallic-like behavior of $%
\delta \sigma _{T}(T)$. Since the interaction in the triplet channel is
enhanced near the QCP, $\delta \sigma _{T}(T)$ is expected to be larger than
$\delta \sigma _{S}(T)$--which is what we find in almost all regimes of
interest. On the other hand, we disregard the weak-localization correction,
which is not relevant for metamagnetic transitions and can readily be
accounted for otherwise. The correction in the triplet channel $\delta
\sigma _{T}(T)$ is calculated within the spin-fermion model of 
Ref.~\cite{chubukov1}. The interaction in this model can be 
treated perturbatively if
$\gamma \gg \alpha$,
where $\gamma $ is the 
dimensionless parameter associated with Landau damping of the spin
fluctuations, and $\alpha $ is the dimensionless coupling between the
electrons and the spin fluctuations. While this relation holds, we are able
to study the various cross-over regimes in the entire $T-\delta $ plane
(where $\delta $ is the distance from the QCP) down to very low temperature.
What is new in our study is that (1) we identify the regime of parameters in
which controlled calculations are possible in the entire $T-\delta $ plane,
(2) we find a new power law dependence ($\delta \sigma \propto -T^{1/3}$) of
the conductivity in the ballistic quantum critical regime, and (3) near the
QCP we find the temperature scale of ballistic-diffusive cross-over to be
much smaller than the nominal scale $1/\tau :$
\begin{equation}
T^{\ast }=1/(\tau  (E_{F}\tau )^{2}\gamma) \ll 1/\tau .  \label{tstar}
\end{equation}

\textit{The model.---} We describe the system by the action
\begin{eqnarray}
S &=&T\sum_{\omega _{n}}\int d^{2}r\:\psi _{\alpha }^{\dagger }(\mathbf{r}%
,\omega _{n})\left( i\omega _{n}+\nabla ^{2}/2m+\mu \right) \psi _{\alpha }(%
\mathbf{r},\omega _{n})  \nonumber  \label{eq:action} \\
&+&(E_{0}T)\sum_{\Omega _{n},\mathbf{q}}U^{-1}(\mathbf{q},\Omega _{n})%
\mathbf{S}(\mathbf{q},\Omega _{n})\cdot \mathbf{S}(-\mathbf{q},-\Omega _{n})
\nonumber \\
&+&\!\!\left( \alpha E_{0}/\nu \right) ^{1/2}\!\!\!\int
\!d^{2}r\!\int_{0}^{\beta }\!\!d\tau \:\psi _{\alpha }^{\dagger }(\mathbf{r}%
,\tau )\psi _{\beta }(\mathbf{r},\tau )\left[ \mathbf{S}(\mathbf{r},\tau
)\cdot \mathbf{\sigma }_{\alpha \beta }\right]  \nonumber \\
&+&\int_{0}^{\beta }d\tau \,\int d^{2}r\,\psi _{\alpha }^{\dagger }(\mathbf{r%
},\tau )V(\mathbf{r})\psi _{\alpha }(\mathbf{r},\tau ),
\end{eqnarray}
where summation over repeated indices is implied. Here ($\psi _{\alpha
}^{\dagger }$, $\psi _{\alpha }$) are Grassman fields for (low-energy)
electrons with spin $\alpha $, $\mathbf{S}(\mathbf{q,}\Omega _{n})$ is a
bosonic field for the collective spin fluctuation modes, $E_{0}$ has
dimension of energy, $\nu =m/\pi $ is the density of states for
non-interacting electrons with spin in 2D, 
and $\mu $ is the chemical potential. Fields $\mathbf{S}(\mathbf{q,}\Omega
_{n})$ are obtained by integrating out electrons above a certain energy
cut-off (for example below which the electron spectrum can be linearized).
The disorder potential $V(\mathbf{r})$ is assumed to obey Gaussian
distribution with $\langle V(\mathbf{r}_{1})V(\mathbf{r}_{2})\rangle =\delta
(\mathbf{r}_{1}-\mathbf{r}_{2})/(2\pi \nu \tau ).$ In our theory the
dimensionless coupling constant $\alpha \lesssim 1$. 

In the ballistic regime the propagator for the spin fluctuations is
\begin{equation}
U(\mathbf{q},\Omega _{n})=\left[ \delta +\left( q/p_{F}\right) ^{2}+\gamma
\left| \Omega _{n}\right| /v_{F}q\right] ^{-1},  \label{eq:chi}
\end{equation}
where $\delta $ is related to the magnetic correlation length $\xi $ by $%
\delta =(p_{F}\xi )^{-2}$. Although the dimensionless parameter $\gamma $ is
not unrelated to the coupling $\alpha $ (for example, $\gamma $ should
vanish when $\alpha $ is zero), the precise relation between the two depends
on microscopic details. In the random phase approximation, $\gamma =\alpha $
~ \cite{chubukov1}. In our theory we take $\gamma $ as an independent
phenomenological parameter.
The
form of the Landau-damping term in Eq.~\eqref{eq:chi} is valid for $%
v_{F}q\gg \Omega $, where it is a universal low-energy feature of itinerant
electrons~ \cite{agd}. In the opposite limit of $\Omega \gg v_{F}q$, the
Landau-damping term depends on microscopic details, and the spin-fermion
model loses universality. We find that in the ballistic regime either $%
v_{F}q\gg \Omega $ (thus justifying the universal form of the Landau
damping), or the contribution of the dynamic term in Eq.~\eqref{eq:chi} is
negligible to leading order. In this sense our results are universal. In the
diffusive limit, the phenomenological form of the spin fluctuation
propagator is given by replacing the dynamic term in Eq.~\eqref{eq:chi} by $%
\gamma |\Omega _{n}|/Dq^{2}$, where $D=v_{F}^{2}\tau /2$ is the diffusion
constant.

\begin{figure}[tbp]
\begin{center}
\includegraphics[width=7cm, height=7cm, trim= 180 480 220 160]{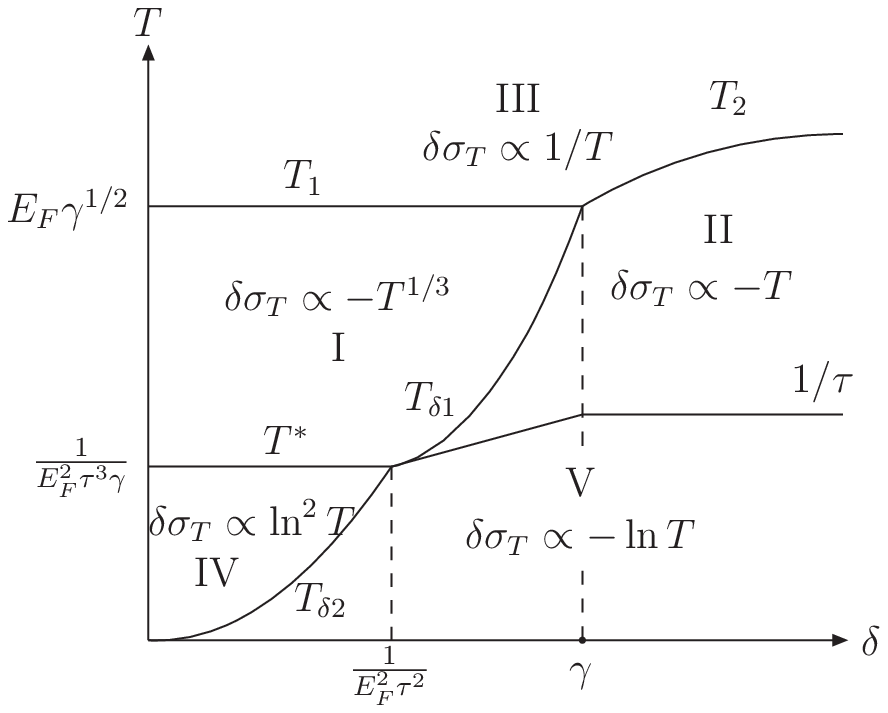}
\end{center}
\caption{Different cross-over regimes for the temperature dependence of
the triplet channel contribution to conductivity.
$T_{\protect\delta 1}=(\protect\delta ^{3/2}/\protect\gamma
)E_{F}$, $T_{\protect\delta 2}=(\protect\delta ^{2}\protect\tau /\protect%
\gamma )E_{F}^{2}$, $T_{2}=E_{F}\protect\delta ^{1/2}$. Notice that $\protect%
\gamma ^{1/2}\gg 1/(E_{F}\protect\tau )$.}
\label{fig:scales}
\end{figure}

Near the QCP, there are two important temperature scales. (i) The
temperature scale $T^{\ast }$~\cite{mirlin} of the
cross-over between ballistic and diffusive motion of the electrons. The
cross-over occurs when the distance travelled by an electron during
interaction, which by uncertainty relation is $1/q$ for momentum transfer $q$%
, is comparable to the distance $v_{F}\tau $ travelled by electrons between
successive impurity scatterings. Very close to the QCP ($\delta \ll \left(
E_{F}\tau \right) ^{-2}$), the momentum transfer $q_{B1}\sim p_{F}(\gamma
\Omega /E_{F})^{1/3}$ is determined by the pole of the propagator in Eq.~%
\eqref{eq:chi}. Since $\Omega \sim T$, we get the cross-over scale $T^{\ast
} $ in Eq. (\ref{tstar}). In the FL-regime far away from the QCP ($\delta
\gg \gamma $), $q$ is of order of the typical momentum of fermionic
excitations $q_{F}\sim \Omega /v_{F}$, and the ballistic-diffusive
cross-over scale is $1/\tau $. In the FL-regime close to the QCP ($\left(
E_{F}\tau \right) ^{-2}\ll \delta \ll \gamma $), $q\sim q_{B2}\sim (\gamma
\Omega )/(v_{F}\delta )$ is still controlled by the pole in 
Eq.~\eqref{eq:chi}, 
and the ballistic-diffusive cross-over scale is $\delta /(\gamma \tau )$.
(ii) $T_1= \tg^{1/2} E_F$ is the scale above which $q_F \gg q_{B1}$,
and the effect of the QCP on the conductivity is small.
We identify two possible situations depending on the strength of disorder
relative to the Landau damping parameter.
(a) For $\gamma^{1/2} > 1/(E_F \tau)$,
the low-temperature cut-off of the regime where $\delta \sigma \propto
-T^{1/3}$ is $T^{\ast }$ and the high-T cut-off 
is $T_1$ (see Fig.~\ref{fig:scales}).
For $T<T^{\ast }$, we recover the result of
Ref.~\cite{kim} with $\delta \sigma _{T}\propto \ln ^{2}(T)$ (however, the
(metallic) sign of our result is opposite to that in Ref.~\cite{kim}).
Above $T_{1}$ the correction in the triplet channel 
$\delta \sigma _{T}\propto 1/T$ is smaller than the singlet-channel one 
and $\delta \sigma \approx \delta \sigma _{S}=e^{2}T\tau /\pi $~\cite{zna}. 
(b) For $1/(E_{F}\tau )>\gamma ^{1/2}$, a
situation which is experimentally highly improbable,
the $T^{1/3}$ regime is lost. 

The  $T^{1/3}$ scaling of $\delta \sigma_T$ in the ballistic quantum 
critical regime in 2D can be simply understood from the following argument.
The correction to the scattering rate due to electron-electron interaction
can be estimated as $\Delta [1/\tau] \sim \left( 1/\tau \right) {\rm Im}
\Sigma \Delta t$, where Im$\Sigma$ is determined by the interaction 
between the electrons mediated by the spin fluctuations, and 
$\Delta t$ is the interaction time. By uncertainty principle, $\Delta t\sim
1/v_{F}q$. In the FL-regime, Im$\Sigma \propto T^{2}$ and 
$q\propto \Omega \sim T$, hence $\Delta [1/\tau] \propto T$.
In 2D
near the QCP, and an interaction with a dynamical exponent 
$z$ (in our case $z=3$ in the ballistic regime, see Eq.~\eqref{eq:chi}),
Im$\Sigma \propto T^{(1-1/z)}$~\cite{rech} and $q\propto \Omega ^{1/z}\propto
T^{1/z}$, hence $\Delta [1/\tau] \propto T^{(1-2/z)}$.

We summarize the technical details~\cite{zna} of the intermediate steps.
First, using Kubo formalism we expand the current-current correlator to the
lowest order in $\alpha $. In the ballistic regime near the QCP, the vertex
correction to the spin-fermion coupling gives contribution which is smaller
by a factor $(\alpha /\gamma ^{1/2})(T/T_{1})^{1/3}\ln (T_{1}/T)$ for $T\ll
T_{1}$, and by a factor $\alpha /\gamma ^{1/2}$ for $T\gg T_{1}$. In the
diffusive regime the next order in coupling $\alpha $ is smaller by $\alpha
\ln ^{2}(T)/(E_{F}\tau )$. As a result, the expansion in the coupling
constant $\alpha $ is controlled. The second step is to perform the analytic
continuation. In the third step we average over disorder.
The correction to the conductivity in the triplet channel can be written as
\cite{zna}
\begin{eqnarray}
\delta \sigma _{T} &=&-(3\pi e^{2}v_{F}^{2}\tau \alpha )\int_{-\infty
}^{\infty }\frac{d\Omega }{4\pi ^{2}}\left[ \frac{\partial }{\partial
\Omega }\left( \Omega \coth \frac{\Omega }{2T}\right) \right]  \nonumber
\label{eq:cond2} \\
&\times &\mathrm{Im}\int \frac{d^{2}q}{(2\pi )^{2}}U^{A}(\mathbf{q},\Omega
)B(\mathbf{q},\Omega ),  \nonumber
\end{eqnarray}
where $B(\mathbf{q},\Omega )$ is the fermionic part of the current-current
correlator (see Eq.~(3.26) in Ref.~\cite{zna}). In the ballistic regime $%
v_{F}q\gg 1/\tau $, and the limiting form of $B$ is given by the term
leading in $\tau $. This is equivalent to an expansion in $(T/T^{\ast
})^{1/3}$ near the QCP, and in $1/(T\tau )$ for $\delta \gg \gamma $. In
this limit $B\approx B_{b}$, where $B_{b}$ is given by
\[
B_{b}(\mathbf{q},\Omega )=(2/(v_{F}q)^{2})(1-(i\Omega
)/S)^{2}+(2/S^{2})(1-(i\Omega )/S),
\]
where $S=((v_{F}q)^{2}-\Omega ^{2}+i\eta \mathrm{Sgn}(\Omega ))^{1/2}$. In
the diffusive regime $v_{F}q\ll 1/\tau $, and the typical momentum is given
by the diffusion pole. In this limit $B\approx B_{d}$, where
\[
B_{d}(\mathbf{q},\Omega )=(\tau (v_{F}q)^{2})/(i\Omega +Dq^{2})^{3}.
\]

\textit{Results.---} The ballistic limit is defined by $T\gg T^{\ast }$ for $%
\delta \ll \left( E_{F}\tau \right) ^{-2}$, by $T\gg \delta /\gamma \tau $
for $\left( E_{F}\tau \right) ^{-2}\ll \delta \ll \gamma $, and by $T\gg
1/\tau $ for $\delta \gg \gamma $. In this limit, there are three cross-over
regimes (regions I-III in Fig.~\ref{fig:scales}).

(1) Regime I. The limiting form of $U$ is obtained by setting $\delta = 0$
in Eq.~\eqref{eq:chi}, which gives the bosonic momentum scale $q_{B1} \sim
p_F (\gamma \Omega/E_F)^{1/3}$. This is the momentum transferred by the spin
fluctuations to the electrons during elastic scattering. In this regime, $%
q_{B1} \gg q_F \sim \Omega/v_F$. Since, $v_F q_{B1} \gg \Omega$, the form of
$B_b$ simplifies to $B_b \approx 4/(v_F q)^2$. The leading temperature
dependence of the conductivity is given by the triplet channel contribution
\begin{equation}  \label{eq:main-result}
\delta \sigma_T (T) = - \frac{e^2 \tau \alpha}{\pi \gamma^{2/3}} \mathcal{C}
(p_F v_F)^{2/3} T^{1/3},
\end{equation}
where $\mathcal{C} = \int_0^{\infty} dt \frac{\partial}{\partial t} \left(
\frac{2t} {1 - e^t} \right) \frac{1}{t^{2/3}} \approx 3.44 $. Eq.~%
\eqref{eq:main-result} is the main result of this Letter. The high
temperature cut-off of this regime is $T_1$, above which fermionic momentum $%
q_F \gg q_{B1}$. At finite $\delta$ the regime ends when $\delta \sim
(q_{B1}/p_F)^2$. This gives the cross-over scale $T_{\delta 1} = E_F
\delta^{3/2}/\gamma$. For temperature below $T_{\delta 1}$ the effect of
finite $\delta$ is important.

(2) Regime II. Two situations can be identified in this regime. For $\delta
\ll \gamma $, the approximate form of $U$ is given by dropping the $%
(q/p_{F})^{2}$-term in Eq.~\eqref{eq:chi}. The dominant momentum scale is $%
q_{B2}\sim (\gamma \Omega )/(v_{F}\delta )\gg q_{F}$, and $B_{b}\approx
4/(v_{F}q)^{2}$. For $\delta \gg \gamma $, the typical momentum scale is
given by $q_{F}\sim \Omega /v_{F}$. In this limit the Landau damping term is
order $\gamma \ll \delta $, and so $U\approx 1/\delta $. For both cases the
triplet channel contribution is
\begin{equation}
\delta \sigma _{T}(T)=-\left( 3e^{2}\tau \alpha /\pi \delta \right) T.
\end{equation}
For $\delta \gg \gamma $, this regime is cut-off at
$T_{2}=\delta^{1/2}E_{F} $, above which $(q/p_{F})^{2}$ term in
$U$ dominates since $(q/p_{F})^{2}\sim (\Omega /E_{F})^{2}\gg \delta $.

(3) Regime III. This is the high temperature regime of the theory where the
typical momentum scale is given by $q_{F}$. For $\delta \ll \gamma $, the
dynamic term in the spin fluctuation propagator can be neglected since $%
(q/p_{F})^{2}\sim (\Omega /E_{F})^{2}\gg \gamma $. For $\delta \gg \gamma $,
the mass of the spin fluctuations can be neglected since $(q/p_{F})^{2}\gg
\delta $. Thus, in this regime $U\approx (p_{F}/q)^{2}$. The leading order
contributions to $\delta \sigma _{T}(T)$ cancel out, and the triplet
channel gives a small contribution to the conductivity $\sigma _{T}\propto
1/T$ . The interference correction is dominated by the contribution from the
singlet channel $\delta \sigma \approx \delta \sigma _{S}\propto T.$ 

In the diffusive limit there are two cross-over regimes (see Fig.~\ref
{fig:scales}).

(1) Regime IV. Setting $\delta =0$, in this regime $U^{-1}(\mathbf{q},\Omega
_{n})\approx ((q/p_{F})^{2}+(\gamma |\Omega _{n}|)/(Dq^{2}))$. The leading
temperature dependence of the conductivity comes from the triplet channel
\begin{equation}
\delta \sigma _{T}(T)=\left( 3/8\pi ^{2}\right) \left( e^{2}\alpha /\gamma
\right) \ln ^{2}\left( \gamma Dp_{F}^{2}/T\right) ,
\end{equation}
which is guaranteed to win over the singlet one [$\delta \sigma
_{S}=-(e^{2}/2\pi ^{2})\ln (E_{F}/T)$] at low enough $T.$ This regime has been
discussed in the context of 2D
metamagnetic QCP~\cite{kim}, and also in the context of fermion gauge field
models~\cite{mirlin,khveshchenko}. 
However, our result leads to a metallic sign of the
conductivity, which was not noticed in prior work~\cite{kim}. 
At finite $\delta $ this
regime exists for $T>T_{\delta 2}=(\delta ^{2}Dp_{F}^{2})/\gamma $.

(2) Regime V. For $T<T_{\delta 2}$, the mass of the spin fluctuations is
important, and $U^{-1}(\mathbf{q},\Omega _{n})\approx (\delta +(\gamma
|\Omega _{n}|)/(Dq^{2}))$. The leading temperature dependence of the
conductivity in the spin channel is
\begin{equation}
\delta \sigma _{T}(T)=\left( 3\mathcal{B}/2\pi ^{2}\right) \left(
e^{2}\alpha /\gamma \right) \ln \left( E_{F}/T\right) ,  \label{spindif}
\end{equation}
where $\mathcal{B}=\ln (\gamma /\delta )$ for $\delta \ll \gamma $, and $%
\mathcal{B}=\gamma /(2\delta )$ for $\delta \gg \gamma $. This is the
Altshuler-Aronov~\cite{altshuler2}
correction to the conductivity for the triplet channel in the diffusive
regime of good metals. 

We now turn to the application of our theory to experimental results. In Sr$%
_{3}$Ru$_{2}$O$_{7}$ the velocity $v_{F}/\gamma $ of the spin fluctuations
is 
presumed to be of the order of Fermi velocity, \emph{i.e.}, $\gamma \sim 1$~
\cite{kim}. Since the in-plane (ab) residual resistivity is 
$\rho \sim $ 2.5 $ \mu \Omega \cdot $cm~\cite{grigera}, 
and the distance between RuO$_{2}$
bilayers is 10 \AA ~\cite{capogna}, the residual resistivity per square is $%
\rho _{2d}\sim $ 25 $\Omega $. Taking $E_{F}\sim 500$ K, we get $
1/\tau \sim $ 4 K. 
By comparing the elastic transport rate due to interaction correction
$1/\tau_{\rm el} \propto {\rm Im} \Sigma/(v_F q \tau) \propto T^{1/3}$, 
with the inelastic transport rate due to interaction with the spin 
fluctuations 
$1/\tau_{\rm in} \propto (q/p_F)^2 {\rm Im} \Sigma \propto T^{4/3}$,
we expect the quantum correction to be important well below a 
temperature scale $1/(\gamma \tau) \sim $ 4 K.  
Experimentally, the resistivity
is observed to follow $T^{r}$ dependence
down to 4 K with $r\approx 1.2$~\cite{perry}. Within our theory, we
understand the exponent $r$
as a competition between lattice-mediated
inelastic processes above $1/(\gamma \tau)$ 
leading to $T^{4/3}$ behavior, and quantum
interference effects dominating below $1/(\gamma \tau)$. We argue that the
experimentally observed exponent may be less than 4/3 due to a pre-cursor
contribution of the $T^{1/3}$ law. Below 4 K the temperature dependence
of conductivity is expected to have the form 
$\delta \sigma (T)=-aT^{1/3}+bT$, 
where the latter is the regular contribution of the singlet 
channel~\cite{zna}. We expect the correction to the conductivity to go
from metallic to insulating behavior below 4 K. This could explain the
dip in the resistivity observed
around 1 K in this compound~\cite{perry}. Notice that the scale 
$1/(\gamma \tau)$ can
be increased by increasing disorder.

\textit{Conclusions.---}
Using the spin-fermion model, we studied the quantum interference 
correction to the conductivity of a 2D disordered itinerant electron 
system close to a ferromagnetic QCP.
Quantum critical fluctuations affect dramatically the temperature
dependence of the conductivity, which  behaves as  
$\delta\sigma\propto  -T ^{1/3}$ and $\delta\sigma\propto \ln^2T$
in the ballistic and diffusive regimes respectively. Near the
QCP the cross-over temperature between ballistic and diffusive dynamics is 
$T^{\ast}=1/(\tau \gamma (E_{F}\tau )^{2})$.
 We estimate that quantum intereference dominates
the $T$-dependence of $\sigma$ for $T \lesssim 1/\tau$.

It is our pleasure to thank A. V. Chubukov, Y. B. Kim, A. J.
Millis and A. P. Mackenzie 
for stimulating discussions. The work of D. L. M. was
supported by NSF Grant No. DMR-0308377. He also acknowledges the
hospitality of the Abdus Salam ICTP.

\end{document}